\documentclass[aip,jcp,reprint,showkeys,noshowkeys]{revtex4-1}

\usepackage{graphicx,dcolumn,bm,xcolor,multirow,amsmath,amssymb,amsfonts,physics,float,lineno,dcolumn,tabularx}
\usepackage[version=4]{mhchem}
\usepackage[normalem]{ulem}
\usepackage{algorithmicx,algorithm,algpseudocode}
\usepackage[tracking]{microtype}
\usepackage{mathpazo,libertine}

\newcommand{\alert}[1]{\textcolor{black}{#1}}
\definecolor{darkgreen}{RGB}{0, 180, 0}

\newcommand{\mc}{\multicolumn}
\newcommand{\mcc}[1]{\multicolumn{1}{c}{#1}}

\newcommand{\FCI}{FCI}
\newcommand{\exFCI}{exFCI}

\newcommand{\sCI}{sCI}
\newcommand{\PT}{PT2}
\newcommand{\Ak}{Ak}
\newcommand{\Bk}{Bk}
\newcommand{\BkO}{Bk$_0$}
\newcommand{\sBk}{s\Bk}
\newcommand{\sBkO}{s\BkO}

\newcommand{\sCIPT}{\sCI-\PT}
\newcommand{\sCIsBk}{\sCI-\sBk}
\newcommand{\sCIsBkO}{\sCI-\sBkO}

\newcommand{\EexFCI}{E_\text{\exFCI}}

\newcommand{\E}[1]{E_{#1}}
\newcommand{\ESS}[2]{E^{(#1)}_{#2}}
\newcommand{\DIaa}{D_{\alpha \alpha}^{(1)}}
\newcommand{\haI}{h_{\alpha I}}
\newcommand{\PsiSS}[2]{\Psi^{(#1)}_{#2}}
\newcommand{\Ndet}{N_\text{det}}
\newcommand{\Nst}{N_\text{st}}
\newcommand{\Ndav}{N_\text{dav}}

\newcommand{\ecSS}[2]{c_{#1}^{(#2)}}

\newcommand{\hH}{\Hat{H}}

\newcommand{\kal}{\ket*{\alpha}}
\newcommand{\kI}{\ket*{I}}

\newcommand{\bH}{\boldsymbol{H}}
\newcommand{\bc}[1]{\boldsymbol{c}_{#1}}

\newcommand{\bDr}[1]{\boldsymbol{\Delta}_{#1}}
\newcommand{\bDrBk}[1]{\boldsymbol{\Delta}^\text{\Bk}_{#1}}
\newcommand{\bDrsBk}[1]{\boldsymbol{\Delta}^\text{\sBk}_{#1}}

\newcommand{\bHSS}[1]{\boldsymbol{H}^{(#1)}}
\newcommand{\bh}{\boldsymbol{h}}
\newcommand{\bg}{\boldsymbol{g}}

\newcommand{\bDSS}[1]{\boldsymbol{D}^{(#1)}}

\newcommand{\bHeffSS}[1]{\boldsymbol{H}^\text{eff}_{#1}}
\newcommand{\bHeffMS}{\boldsymbol{H}^\text{eff}}

\newcommand{\bcSS}[2]{\boldsymbol{c}^{(#1)}_{#2}}

\newcommand{\bDrSS}[1]{\boldsymbol{\Delta}^\text{sBk}}
\newcommand{\bDrMS}{\boldsymbol{\Delta}^\text{sBk}}
\newcommand{\bdrMS}{\boldsymbol{\delta}^\text{sBk}}
\newcommand{\bDrHe}{\Tilde{\boldsymbol{\Delta}}^\text{sBk}}

\newcommand{\bcMS}[1]{{\boldsymbol{c}^{(#1)}}}
\newcommand{\bS}{\boldsymbol{S}}
\newcommand{\bs}{\boldsymbol{s}}

\newcommand{\bU}{\boldsymbol{U}}

\newcommand{\bE}{\boldsymbol{E}}

\newcommand{\bI}{\boldsymbol{1}}
\newcommand{\bO}{\boldsymbol{0}}

\newcommand{\spp}{$\sigma$+$\pi$}

\renewcommand{\tr}[1]{{}^{\dag}{#1}}
\newcommand{\SI}{supplementary material}

\newcommand{\LCPQ}{Laboratoire de Chimie et Physique Quantiques, Universit\'e de Toulouse, CNRS, UPS, France}
\newcommand{\LCT}{Laboratoire de Chimie Th\'eorique, Universit\'e Pierre et Marie Curie, Sorbonne Universit\'e, CNRS, Paris, France}

\begin{document}	

\title{\textcolor{blue}{\textbf{Selected configuration interaction dressed by perturbation}}}

\author{Yann Garniron}
\affiliation{\LCPQ}
\author{Anthony Scemama}
\affiliation{\LCPQ}
\author{Emmanuel Giner}
\affiliation{\LCT}
\author{Michel Caffarel}
\affiliation{\LCPQ}
\author{Pierre-Fran{\c c}ois Loos}
\email[Corresponding author: ]{loos@irsamc.ups-tlse.fr}
\affiliation{\LCPQ}

\begin{abstract}
Selected configuration interaction (sCI) methods including second-order perturbative 
corrections provide near full CI (FCI) quality energies with only a small fraction 
of the determinants of the FCI space. 
Here, we introduce both a state-specific and a multi-state sCI method based on the CIPSI (Configuration Interaction using a Perturbative Selection made Iteratively) algorithm.
The present method revises the reference (internal) space under the effect of its interaction with the outer space via the construction of an effective Hamiltonian, following the shifted-Bk philosophy of Davidson and coworkers.
In particular, the multi-state algorithm removes the storage bottleneck of the effective Hamiltonian via a low-rank factorization of the dressing matrix.
Illustrative examples are reported for the state-specific and multi-state versions.
\end{abstract}

\maketitle

\section{Introduction}
Recently, selected configuration interaction (sCI) methods have 
demonstrated their ability to reach, \alert{for moderate size basis sets}, near full CI (FCI) quality energies for small organic and transition metal-containing molecules. \cite{Giner_2013, Caffarel_2014, Giner_2015a, Garniron_2017, Caffarel_2016, Caffarel_2016b, Holmes_2016, Sharma_2017, Holmes_2017, Chien_2018, Scemama_2018a, Loos_2018b, Scemama_2018b}
\alert{Selecting iteratively the most relevant determinants of the FCI space is an old idea that, to the best of our knowledge, dates back to the pioneering works of Bender and Davidson, \cite{Bender_1969} and Whitten and Hackmeyer \cite{Whitten_1969} in 1969.
Few years later, Huron et al.~\cite{Huron_1973} proposed the so-called CIPSI (Configuration Interaction using a Perturbative Selection made Iteratively) approach to complement the variational sCI energy with a second-order Epstein-Nesbet perturbative correction. 
This has demonstrated to be a particularly efficient way of approaching the FCI limit. \cite{Garniron_2017b, Sharma_2017, Blunt_2018, Loos_2018b, Scemama_2018b, Scemama_2018a}
Over these last few years, we have witnessed a resurgence of sCI methods under various variants and acronyms. 
In short, their main differences lie in the way i) the determinant selection is done and, ii) the second-order contribution is computed. 
The selection can be done purely stochastically as in FCIQMC \cite{Booth_2009} or deterministically as in CIPSI or other variants, such as heat-bath CI, \cite{Holmes_2016, Sharma_2017, Holmes_2017, Chien_2018} adaptive sampling CI (ASCI) \cite{Evangelista_2014, Schriber_2016, Tubman_2016} or iterative CI (ICI). \cite{Liu_2016} 
Similarly, the second-order correction can be computed either purely deterministically or semi-stochastically by a Monte Carlo sampling. \cite{Garniron_2017, Sharma_2017, Blunt_2018}
Here, we shall use the CIPSI method \cite{Huron_1973} to generate the model space, but any other sCI variants could be employed.}

For a given electronic state $k$, the ensemble of determinants $\kI$, which constitutes the zeroth-order (normalized) wave function
\begin{equation}
\label{eq:Psi0}
	\ket*{\PsiSS{0}{k}} =  \sum_{I=1}^{\Ndet} \ecSS{Ik}{0} \kI
\end{equation}
of (variational) zeroth-order energy 
\begin{equation}
\label{eq:EO}
	\ESS{0}{k} 
	= \mel*{\PsiSS{0}{k}}{\hH}{\PsiSS{0}{k}}
	= \tr{\bcSS{0}{k}} \bHSS{0} \bcSS{0}{k},
\end{equation}
(where $\tr{\bcSS{0}{k}}$ are the transposed coefficients) defines the (zeroth-order) reference model space, or internal space.
The remaining determinants of the FCI space belong to the external space, or outer space.
In particular, the ensemble of determinants $\kal$ connected to $\PsiSS{0}{k}$, i.e.,  $\mel*{\alpha}{\hH}{\PsiSS{0}{k}} \neq 0$ and $\braket*{\alpha}{\PsiSS{0}{k}} = 0$ --- the so-called ``perturbers'' --- defines the (first-order) perturbative space, such as
\begin{align}
\label{eq:Psi1}
	\ket*{\PsiSS{1}{k}} & =  \sum_{\alpha} \ecSS{\alpha k}{1} \kal,
	&
	\bcSS{1}{k} & =  (\ESS{0}{k} \bI - \bDSS{1})^{-1} \bh \bcSS{0}{k}, 
\end{align}
where $\bI$ is the identity matrix, $\bDSS{1}$ is a diagonal matrix with elements $\DIaa = \mel*{\alpha}{\hH}{\alpha}$ and $\haI = \mel*{\alpha}{\hH}{I}$.
Within CIPSI, the ``distance'' to the FCI solution is estimated via a second-order Epstein-Nesbet perturbative energy correction:
\begin{equation}
\label{eq:EPT2}
	\ESS{2}{k} 
	=  \mel*{\PsiSS{0}{k}}{\hH}{\PsiSS{1}{k}} 
	= \tr{\bcSS{0}{k}}\ \tr{\bh}\ \bcSS{1}{k}.
\end{equation}

The second-order correction \eqref{eq:EPT2} has obvious advantages and can be computed efficiently using diagrammatic\cite{Cimiraglia_1996} or hybrid stochastic-deterministic approaches. \cite{Garniron_2017b, Sharma_2017, Blunt_2018}
However, it has also an obvious disadvantage: the internal space is not revised under the effect of its interaction with the outer space. 
Here, thanks to intermediate effective Hamiltonian theory, \cite{Malrieu_1985} we propose to build and diagonalize an effective Hamiltonian taking into account the effect of the perturbative space. \cite{Giner_2017a, Pathak_2017}
\alert{This idea is based on the so-called Bk method, originally proposed by Gershgorn and Shavitt \cite{Gershgorn_1968} and later refined and rebranded shifted-Bk (\sBk) by Davidson and coworkers. \cite{Nitzsche_1978a, Nitzsche_1978b, Davidson_1981, Rawlings_1983, Rawlings_1984, Kozlowski_1994a, Kozlowski_1994b, Kozlowski_1994c, Kozlowski_1995, Staroverov_1998}
(See also Refs.~\onlinecite{Nakano_1993, Nakano_2001, Kirtman_1981, LiManni_2013}.)
All these works lie on the seminal idea of L\"owdin on the partition of the FCI Hamiltonian matrix. \cite{Lowdin_1951}
Initially, Gershgorn and Shavitt \cite{Gershgorn_1968} introduced several approximations, two of them being denoted {\Ak} and {\Bk}. 
Both use a partitioning of the CI matrix based on the selection of a dominant subset of (primary) configurations.
The {\Ak} method, which is related to earlier work by Claverie, Diner and Malrieu, \cite{Claverie_1967} estimates the contribution of the configurations left out of the CI expansion, an idea very similar to the computation of the second-order correction [see Eq.~\eqref{eq:EPT2}]. \cite{Bender_1969, Buenker_1975}
Compared to the {\Ak} method, the coefficients of the primary configurations are allowed to relax in the {\Bk} method.
The different flavours of {\Bk} methods are usually due to the distinct partition of the Hamiltonian matrix, and the reference energy used to define the perturbers [see Eq.~\eqref{eq:Psi1} and discussion below]. \cite{Nitzsche_1978a, Nitzsche_1978b, Davidson_1981, Rawlings_1983, Rawlings_1984, Kozlowski_1994a, Kozlowski_1994b, Kozlowski_1994c, Kozlowski_1995, Staroverov_1998, Nakano_1993, Nakano_2001, Kirtman_1981, LiManni_2013, Giner_2017a, Pathak_2017}}

To be best of our knowledge, the shifted-{\Bk} method has never been coupled with CIPSI-like sCI methods.
Moreover, in addition to its convergence acceleration to the FCI limit, one of the interesting advantage of shifted-{\Bk} is to provide an explicit revised wave function that one can use, for example, as a trial wave function within quantum Monte Carlo. \cite{Giner_2013, Caffarel_2014, Caffarel_2016, Caffarel_2016b, Scemama_2018a, Scemama_2018b} 
In the present manuscript, we propose both a state-specific and a multi-state formulation which remove the storage bottleneck of the effective Hamiltonian.
Furthermore, the present computations are performed semi-stochastically as in our recently proposed hybrid stochastic-deterministic algorithm for the computation of $\ESS{2}{}$. \cite{Garniron_2017b}
Unless otherwise stated, atomic units are used throughout.

\section{Shifted-Bk
\label{sec:sBk}
}

\subsection{State-specific shifted-Bk
\label{sec:SS}}

For a given electronic state $k$, in order to solve the Schr{\"o}dinger equation \alert{$\bH \bc{k} = \E{k} \bc{k}$} in the FCI space, the eigenvalue problem may be partitioned as
\begin{equation}
\label{eq:CI-partition}
	\begin{pmatrix}
		\bHSS{0}	&	\tr{\bh}	&	\bO			\\
		\bh			&	\bHSS{1}	&	\tr{\bg}	\\
		\bO			&	\bg			&	\bHSS{2}	\\
	\end{pmatrix}
	\begin{pmatrix}
		\bcSS{0}{k}	\\
		\bcSS{1}{k}	\\
		\bcSS{2}{k}	\\
	\end{pmatrix}
	- \E{k}
	\begin{pmatrix}
		\bcSS{0}{k}	\\
		\bcSS{1}{k}	\\
		\bcSS{2}{k}	\\
	\end{pmatrix}
	= 
	\begin{pmatrix}
		\bO	\\
		\bO	\\
		\bO	\\
	\end{pmatrix},
\end{equation}
where $\bHSS{2}$ is the second-order Hamiltonian corresponding to the external configurations excluding the perturbers, and $\bg$ is the coupling matrix between first- and second-order spaces.
Equation \eqref{eq:CI-partition} can be recast as an ``effective'' Schr{\"o}dinger equation $\bHeffSS{k} \bcSS{0}{k} = \E{k} \bcSS{0}{k}$ with the effective Hamiltonian 
\begin{equation}
	\label{eq:Heff}
	\bHeffSS{k} = \bHSS{0} + \bDr{k},
\end{equation}
and dressing matrix
\begin{equation}
	\label{eq:Dr}
	\bDr{k} =  \tr{\bh} \qty[ (\E{k} \bI - \bHSS{1}) - \tr{\bg} (\E{k} \bI - \bHSS{2})^{-1} \bg]^{-1} \bh.
\end{equation}
Within the state-specific version of the {\Bk} method introduced by Gershgorn and Shavitt, \cite{Gershgorn_1968} for each target electronic state $k$, we i) approximate $\bHSS{1}$ by its (diagonal) zeroth-order approximation $\bDSS{1}$, and ii) neglect the influence of the second-order space $\bHSS{2}$.
Hence, the state-specific {\Bk} dressing matrix is defined as
\begin{equation}
	\label{eq:DrBk}
	\bDrBk{k} = \tr{\bh} (\E{k} \bI - \bDSS{1})^{-1} \bh,
\end{equation}
which naturally yields to a Brillouin-Wigner perturbation approximation. \cite{Gershgorn_1968}

The shifted-{\Bk} method of Davidson and coworkers \cite{Nitzsche_1978a, Nitzsche_1978b, Davidson_1981, Rawlings_1983, Rawlings_1984} still approximates $\bHSS{1}$ by its diagonal $\bDSS{1}$, but \textit{``shifts''} (hence the name) the energy at the denominator of Eq.~\eqref{eq:Dr} to take into account the influence of the second-order term $\tr{\bg} (\E{k} \bI - \bHSS{2})^{-1} \bg$, in other words
\begin{equation}
	\label{eq:shift}
	\E{k} \bI - \tr{\bg} (\E{k} \bI - \bHSS{2})^{-1} \bg \approx \ESS{0}{k} \bI.
\end{equation}
Therefore, the state-specific shifted-{\Bk} dressing matrix is 
\begin{equation}
	\label{eq:DrsBk}
	\bDrsBk{k} = \tr{\bh} (\ESS{0}{k} \bI - \bDSS{1})^{-1} \bh,
\end{equation}
which leads to the Epstein-Nesbet variant of Rayleigh-Schr{\"o}dinger perturbation theory. \cite{Davidson_1981, Rawlings_1983}.
Compared to the {\Bk} method, its shifted variant has the indisputable advantage of correcting some of the size-consistency error. \cite{Davidson_1981}
However, as expected, the present methodology is only nearly size-consistent.
Note that the shifted-{\Bk} method is an iterative method as, thanks to the influence of the entire external space, both the zeroth-order coefficients $\bcSS{0}{k}$ and energy $\ESS{0}{k}$ (given by Eq.~\eqref{eq:EO}) are revised at each iteration.

For small CI expansions, it is possible to store the entire dressed Hamiltonian matrix $\bHeffSS{k}$ of size $\Ndet \times \Ndet$.
However, when the CI expansion gets large, $\bHeffSS{k}$ becomes too large to be stored in memory.
Thankfully, it is not necessary to explicitly build $\bHeffSS{k}$.
Indeed, for large CI expansions, we switch to a Davidson diagonalization procedure \cite{Davidson_1975} which only requires the computation of the vectors $\bHSS{0} \bcSS{0}{k}$ and $\bDrsBk{k} \bcSS{0}{k}$ of size $\Ndet$.

\subsection{Multi-state shifted-Bk
\label{sec:MS}
}

In a multi-state calculation, one has to adopt a different strategy in order to dress the Hamiltonian for all the target states simultaneously.
This is particularly important in practice, for instance, to determine accurate vertical transition energies.
An unbalanced treatment of the ground and excited states, even for states with different spatial or spin symmetries, could have significant effects on the accuracy of these energy differences. \cite{Loos_2018b}

For sake of simplicity, let us assume that our aim is to calculate the dressed energy of the $\Nst$ lowest electronic states.
For $1 \le k \le \Nst$, we wish to find a multi-state effective Hamiltonian $\bHeffMS$ and a dressing matrix $\bDrMS$, with $\bHeffMS = \bHSS{0} + \bDrMS$, such that, when applied to the $k$-th state coefficient vector $\bcSS{0}{k}$, one recovers the $k$-th state-specific dressing matrix $\bDrsBk{k}$ times the same vector $\bcSS{0}{k}$, i.e.,
\begin{equation}
	\bDrMS\, \bcSS{0}{k} = \bDrsBk{k}\, \bcSS{0}{k}.
\label{eq:dress_multi}
\end{equation}
\alert{A solution obeying Eq.~\eqref{eq:dress_multi} is
\begin{equation}
        \bDrMS\,  = \sum_{kl} \bDrsBk{k} \bcSS{0}{k} (\bS^{-1})_{kl} \tr{\bcSS{0}{l}},
\end{equation}
where $(\bS^{-1})_{kl} =  \braket*{\bcSS{0}{k}}{\bcSS{0}{l}}$.
In contrast to the state-specific case, $\bHeffMS$ is non-Hermitian as a consequence of the non-orthogonality of the exact state projections 
on the model space. \cite{Malrieu_1985}
In practice, we have found that a robust algorithm can be defined by symmetrizing the multi-state dressing matrix as
\begin{equation}
        \bDrHe =\qty(\tr{\bDrMS} + \bDrMS)/2.
\end{equation}
The eigenstates being now orthonormal, the dressing matrix reduces to
\begin{equation}
\label{eq:DrMS}
  \bDrMS = \bdrMS \, \tr{\bcMS{0}},
\end{equation}
which is reminiscent of a low-rank factorization.
Here, 
\begin{subequations}
\begin{gather}
	\label{eq:cMS}
	\bcMS{0}  = \qty[ \bcSS{0}{1}, \dots, \bcSS{0}{\Nst}],
	\\
	\label{eq:drMS}
	\bdrMS = \qty[ \bDrsBk{1} \bcSS{0}{1}, \dots,  \bDrsBk{\Nst} \bcSS{0}{\Nst}]
\end{gather}
\end{subequations}
are both of size $\Ndet \times \Nst$.}

Two key remarks are in order here:
i) at first order, the symmetrization error is strictly zero, i.e., $\tr{\bcSS{0}{k}} (\bDrMS - \bDrHe) \bcSS{0}{k}  = 0$, and
ii) the symmetrization error becomes vanishingly small for large CI expansions.
Consequently, the symmetrization error can be safely neglected in practice.
\alert{Our preliminary tests have corroborated these theoretical justifications.}
Also, it can be further estimated via second-order perturbation theory.
However, it requires the energies and coefficients of the entire internal space which is only possible for relatively small CI expansions.

The energies of the first $\Nst$ states, $\bE = \qty(\E{1},\ldots,\E{\Nst})$, are obtained by a Davidson diagonalization of the multi-state effective Hamiltonian $\bHeffMS = \bHSS{0} + \bDrHe$.
Similarly to the state-specific case, technically, one is able to store the vectors $\bdrMS$ and $\bcMS{0}$ but $\bDrHe$ (or $\bDrMS$) is potentially too large to be stored in memory.
Luckily, compared to a standard CI calculation, the Davidson diagonalization procedure only requires, at each iteration, the extra knowledge of 
\begin{equation}
\label{eq:extraDr}
	\bDrHe \bU = \qty( \bcMS{0}\ \tr{\bdrMS}\ \bU + \bdrMS \ \tr{\bcMS{0}} \bU )/2.
\end{equation}
where $\bU$ is a $\Ndet \times \Ndav$ matrix gathering the $\Ndav$ vectors considered in the Davidson diagonalization algorithm at a given iteration (with $\Nst \le \Ndav \ll \Ndet$).
Thanks to Eq.~\eqref{eq:DrMS}, this term can be efficiently evaluated in a $\order{\Ndet}$ computational cost and storage via two successive matrix multiplications, for instance,
\begin{gather*}
	\bcMS{0}\ \tr{\bdrMS}\ \bU = \qty[ \bcMS{0} \times \qty( \tr{\bdrMS} \times \bU ) ].
\end{gather*}
A pseudo-code of our iterative multi-state dressing algorithm is presented in {\SI}.
For $\Nst=1$, the present multi-state algorithm reduces to the state-specific version.

\begin{figure}
	\includegraphics[width=0.9\linewidth]{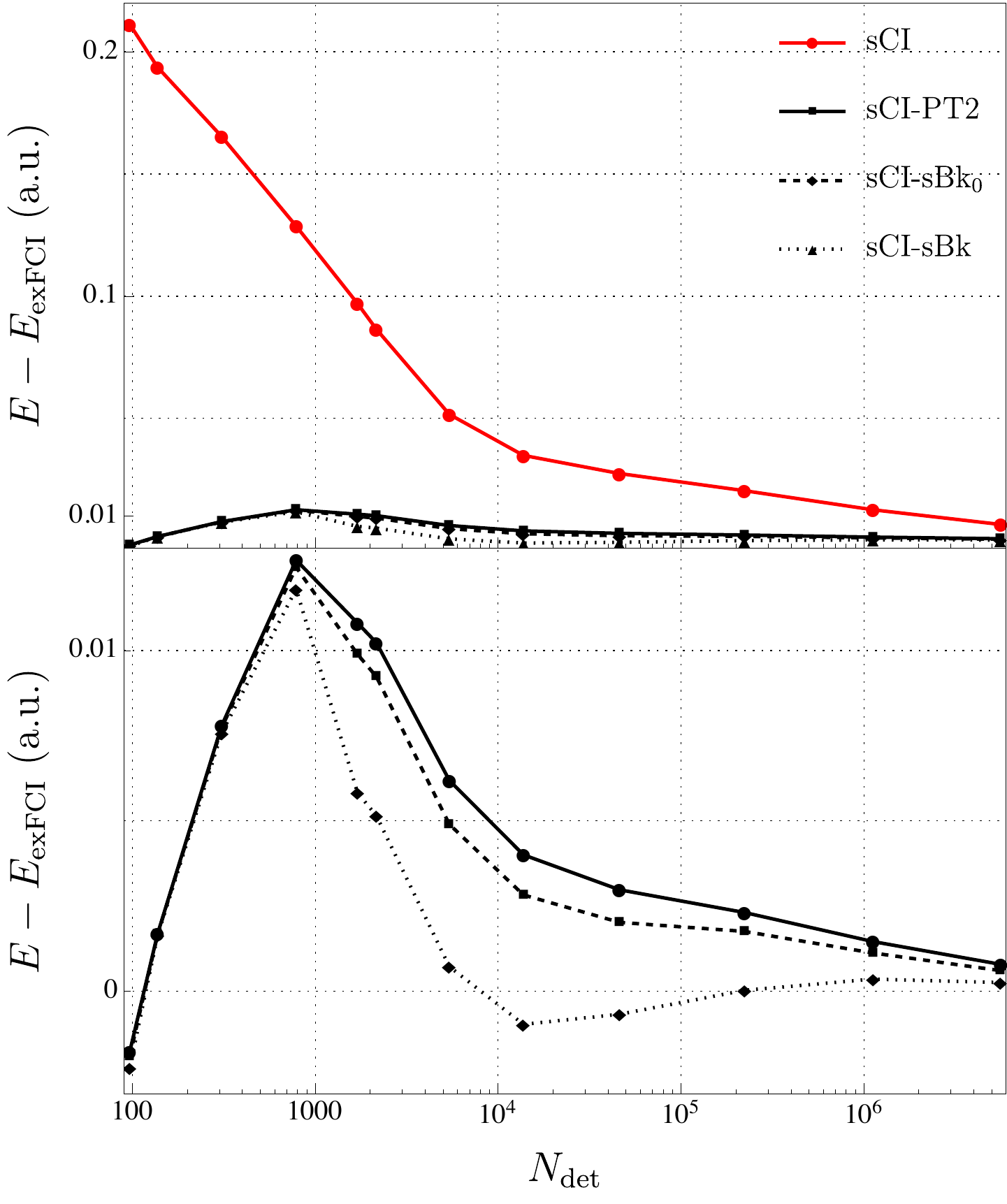}
	\caption{
	\label{fig:CuCl2}
	Deviation from the extrapolated FCI energy $\EexFCI$ of the total energy $E$ of \ce{CuCl2} (in Hartree) as a function of the number of determinants $\Ndet$ in the sCI wave function for various methods.}
\end{figure}

\begin{table*}
\caption{\label{tab:CuCl2}
\alert{
Deviation (in millihartree) from the extrapolated FCI energy ($\EexFCI = -2558.006\,880$ a.u.) for various methods as a function of the number of determinants $\Ndet$ in the CIPSI expansion for the \ce{CuCl2} molecule and the 6-31G basis set.
The second-order correction $\ESS{2}{}$ is also reported.
The error bar corresponding to one standard deviation is reported in parenthesis.
The exFCI energy has been obtained via a linear extrapolation using the energies of the two largest wave functions (see {\SI}).
The two rightmost columns report the overlap with respect to the largest sCI wave function.
}
}
\begin{ruledtabular}
\begin{tabular}{rdddddd}
$\Ndet$			&	\mcc{$\ESS{2}{}$}	&	\mc{3}{c}{$\Delta E$}		&	\mc{2}{c}{Overlap}	\\
									\cline{3-5}						\cline{6-7}
				&					&	\mcc{\sCIPT}				&	\mcc{\sCIsBkO}	&	\mcc{\sCIsBk}	
									&	\mcc{\sCI}				&	\mcc{\sCIsBk}	\\
\hline
$97$			&	-213.039(0)		&	-1.778(0)		&	-1.93(0)	&	-2.25(0)	&	0.9275	&	0.9275		\\
$138$			&	-191.914(0)		&	+1.698(0)		&	+1.68(0)	&	+1.65(0)	&	0.9295	&	0.9295		\\ 
$309$			&	-157.491(0)		&	+7.799(0)		&	+7.74(0)	&	+7.59(0)	&	0.9345	&	0.9345		\\
$789$			&	-116.025(0)		&	+12.654(0)		&	+12.45(0)	&	+11.81(0)	&	0.9438	&	0.9447		\\
$1\,708$		&	-86.208(2)		&	+10.807(2)		&	+9.89(0)	&	+5.83(0)	&	0.9579	&	0.9671		\\
$2\,167$		&	-76.249(8)		&	+10.232(8)		&	+9.23(1)	&	+5.15(1)	&	0.9610	&	0.9700		\\
$5\,428$		&	-45.49(3)		&	+6.19(3)		&	+4.90(3)	&	+0.72(3)	&	0.9777	&	0.9854		\\
$13\,803$		&	-30.87(9)		&	+4.00(9)		&	+2.83(9)	&	-0.97(9)	&	0.9853	&	0.9912		\\
$46\,327$		&	-24.48(9)		&	+2.98(9)		&	+2.02(9)	&	-0.68(9)	&	0.9913	&	0.9952		\\
$223\,089$		&	-18.13(9)		&	+2.31(9)		&	+1.76(9)	&	+0.03(9)	&	0.9956	&	0.9975		\\
$1\,125\,547$	&	-11.18(9)		&	+1.46(9)		&	+1.12(9)	&	+0.36(9)	&	0.9984	&	0.9990		\\
$5\,615\,264$	&	-5.84(2)		&	+0.79(2)		&	+0.61(2)	&	+0.26(2)	&	0.9996	&	0.9997		\\
$26\,493\,179$	&	-3.34(2)		&	+0.45(2)		&	\mcc{---}	&	\mcc{---}	&	1.0000	&	\mcc{---}		\\
\end{tabular}
\end{ruledtabular}
\end{table*}

\section{
Hybrid Stochastic/Deterministic dressings
\label{sec:hybrid}
}

\alert{In Ref.~\onlinecite{Garniron_2017b}, we proposed to express 
\begin{equation}
\label{eq:E2-I}
  \ESS{2}{}  = \sum_{I=1}^{\Ndet} \ESS{2}{[I]}
\end{equation}
as a sum of $\Ndet$ contributions $\ESS{2}{[I]}$, each of them associated with a determinant of the model space, and to compute it efficiently via a Monte Carlo (MC) algorithm.
Thanks to the relatively small size of the MC space ($\Ndet$), one is able to store each single contribution. 
Hence, during the MC simulation, if the contribution of a determinant is required and has never been computed previously, it is computed and stored. 
Otherwise, the value is retrieved from memory. 
This technique, known as \textit{memoization}, drastically accelerates the MC calculation as each contribution needs to be computed only once.
Moreover, we decompose the energy into a deterministic part and a stochastic part, making the deterministic part grows along the calculation until one reaches the desired accuracy.
If desired, the calculation can be carried on until the stochastic part entirely vanishes. 
In that case, the exact result is obtained with no error bar and no noticeable computational overhead compared to the fully deterministic calculation.
To summarize, this algorithm allows to compute a truncated sum with no bias, but with a statistical error bar instead.}

\alert{This algorithm is very general and is not limited to the calculation of $\ESS{2}{}$. 
Similarly to Eq.~\eqref{eq:E2-I}, we express the dressing matrix \eqref{eq:DrMS} as the sum of dressing matrices
\begin{equation}
  \bDrMS  = \sum_{I=1}^{\Ndet} \bDrMS_{[I]}.
\end{equation}
Because the matrices $\bDrMS_{[I]}$ are too large to fit in memory, we sample the vectors $\bdrMS_{[I]}$ [see Eq.~\eqref{eq:drMS}], which are required for the Davidson diagonalization.
During the sampling, one can monitor the ``dressed'' energy as
\begin{equation}
	\E{k} 
	= \mel*{\PsiSS{0}{k}}{\bHeffSS{k}}{\PsiSS{0}{k}} 
	= \ESS{0}{k} + \tr{\bcMS{0}} \expval*{\bdrMS},
\end{equation}
as well as its accuracy by computing the corresponding statistical error. 
In the next section, all {\sBk} calculations have been carried on until the statistical error is below $10^{-5}$ a.u.
Let us emphasize once again that the primary purpose of the present MC algorithm is to accelerate the computation of the dressing matrix.
The same results would have been obtained via its deterministic version.}

\section{Illustrative calculations
\label{sec:example}
}
Unless otherwise stated, all the calculations presented here have been performed with the electronic structure software \textsc{quantum package}, \cite{QP} developed in our group and freely available.
The sCI wave functions are generated with the CIPSI algorithm, as described in Refs.~\onlinecite{Giner_2013, Giner_2015a} in the frozen-core approximation.
The extrapolated {\FCI} results, labeled as exFCI, have been obtained via the method recently proposed by Holmes, Umrigar and Sharma \cite{Holmes_2017} in the context of the heat-bath method. \cite{Holmes_2016, Sharma_2017, Holmes_2017} 
This method has been shown to be robust even for challenging chemical situations, \cite{Scemama_2018a, Chien_2018, Loos_2018b, Scemama_2018b} and we refer the interested readers to Ref.~\onlinecite{Scemama_2018a} for additional details.

\subsection{State-specific example
\label{sec:SS-example}
}
To illustrate the improvement brought by the shifted-{\Bk} approach in its state-specific version (see Sec.~\ref{sec:SS}), we have computed the total electronic energy of the ${}^2\Pi_\text{g}$ ground state of \ce{CuCl2} with the 6-31G basis set. 
The geometry has been taken from Ref.~\onlinecite{Caffarel_2014} where additional information can be found on this system.
For this particular example, we have chosen a small basis set in order to be able to easily reach the FCI limit.
A larger basis set will be considered in the next (multi-state) example (see Sec.~\ref{sec:MS-example}).
The molecular orbitals have been obtained at the restricted open-shell Hartree-Fock (ROHF) level, and the 15 lowest doubly occupied orbitals have been frozen. 
This corresponds to a sCI calculation of 33 electrons in 38 orbitals.
{\sCIPT} stands for a sCI calculation where we have added to the (zeroth-order) variational energy $\ESS{0}{}$ defined in Eq.~\eqref{eq:EO} the value of the second-order correction $\ESS{2}{}$ given by Eq.~\eqref{eq:EPT2}.
The one-shot, non-iterative shifted-Bk procedure will be labeled as {\sCIsBkO}, while its self-consistent version is simply labeled {\sCIsBk}. 

Figure \ref{fig:CuCl2} shows the convergence of the total energy of \ce{CuCl2} as a function of the number of determinants $\Ndet$ in the sCI wave function for \alert{the variational {\sCI} results, as well as} {\sCIPT}, {\sCIsBkO} and {\sCIsBk}. 
The corresponding numerical values are reported in \alert{Table \ref{tab:CuCl2}}.
As expected, the {\sCIPT}, {\sCIsBkO} and {\sCIsBk} energies are not variational as perturbative energies and energies obtained by projection are not guaranteed to be an upper bound of the FCI energy.
\alert{Nonetheless, all of these corrections drastically improve the rate of convergence compared to the variational {\sCI} results (note the logarithmic scale in Fig.~\ref{fig:CuCl2}).}
As shown in the bottom graph of Fig.~\ref{fig:CuCl2}, for small values of $\Ndet$, the three methods yield very similar total energies.
However, for $\Ndet \gtrsim 10^3$, results start to deviate due to the inclusion of an important configuration corresponding to a ligand-to-metal charge transfer (LMCT) state. \cite{Giner_2015b}
This LMCT configuration induces a strong revision of the model space wave function $\PsiSS{0}{}$.
\alert{Because the LMCT configuration corresponds to a singly-excited determinant with respect to the ROHF determinant, it is not included in the CIPSI expansion for small $\Ndet$ values as it does not directly interact with the ROHF reference (Brillouin's theorem).
Therefore, the double excitations which are strongly coupled with the ROHF configuration are first selected by the CIPSI algorithm.
Then, the LMCT configuration is included via its connection with the doubles. 
In particular, the double excitations corresponding to a single excitation on top of the LMCT configuration have been found to strongly interact with it.}
The key observation here is that the {\sCIsBk} energy converges much faster to the FCI limit than the {\sCIPT} energy.
Moreover, the significant difference between {\sCIsBk} and {\sCIsBkO} highlights the importance of the revision of the internal wave function brought by the self-consistent nature of the shifted-Bk method.

\alert{Table \ref{tab:CuCl2} also reports the overlap of the {\sCI} and {\sCIsBk} wave functions with respect to the largest sCI wave function obtained for $\Ndet = 26\,493\,179$.
These results also highlight the faster convergence of {\sCIsBk} and illustrate that the shifted-Bk method could potentially provide better quality trial wave functions for quantum Monte Carlo. \cite{Giner_2013, Caffarel_2014, Caffarel_2016, Caffarel_2016b, Scemama_2018a, Scemama_2018b}}

\alert{Although $\PsiSS{0}{k}$ may be an eigenfunction of $\widehat{S}^2$, the way $\PsiSS{1}{k}$ is built does not enforce this property.
The expectation value of $\widehat{S}^2$ can be monitored by
\begin{equation}
	\label{eq:S2}
	\mel*{\PsiSS{0}{k}}{\widehat{S}^2}{\PsiSS{1}{k}} 
	= \tr{\bcSS{0}{k}}\ \tr{(\bs^2)}\ \bcSS{1}{k}.
\end{equation}
As expected the deviation from the eigenvalue is always small, with a maximum deviation of the order of $10^{-4}$ a.u.~in the case of \ce{CuCl2}.}

\begin{table}
	\caption{
	\label{tab:cya}
	Vertical excitation energy (in eV) of cyanines for various methods.
	The error bar corresponding to one standard deviation is reported in parenthesis.}
	\begin{ruledtabular}
	\begin{tabular}{lddc}
		Method										&	\mcc{CN3}		&	\mcc{CN5}		&	Ref.		\\
		\hline
		CAS($\pi$)\footnotemark[1]	&	7.62	&	5.27	&		this work	\\
		CAS($\pi$)+\PT				&	7.43	&	5.02	&		this work	\\
		CAS($\pi$)+\sBkO			&	7.40	&	4.98	&		this work	\\
		CAS($\pi$)+\sBk				&	7.17	&	4.77	&		this work	\\
		exFCI(\spp)\footnotemark[2]		&	7.17	&	4.89		&		this work	\\
		\hline
		CASSCF($\pi$)\footnotemark[3]		&	7.59		&	5.25	&	Ref.~\onlinecite{Send_2011}	\\
		CASPT2($\pi$)\footnotemark[4]		&	7.26		&	4.74	&	Ref.~\onlinecite{Send_2011}	\\
		CC3(\spp)\footnotemark[5]		&	7.27		&	4.89	&	Ref.~\onlinecite{Send_2011}	\\
		\hline
		DMC\footnotemark[6]			&	7.38(2)		&	5.03(2)	&	Ref.~\onlinecite{Send_2011}	\\
		exCC3(\spp)\footnotemark[7]		&	7.16		&	4.84	&	Ref.~\onlinecite{Send_2011}	\\
	\end{tabular}		
	\end{ruledtabular}
	\footnotetext[1]{CAS-CI/aug-cc-pVDZ calculations: CAS(4,32) and CAS(6,50) for CN3 and CN5, respectively.}
	\footnotetext[2]{Extrapolated CIPSI/aug-cc-pVDZ calculations (see {\SI}).}
	\footnotetext[3]{CASSCF/ANO-L-VDZP calculations with optimal active spaces: CAS(4,6) and CAS(6,10) for CN3 and CN5, respectively.}
	\footnotetext[4]{CASPT2/ANO-L-VDZP calculations with the standard IPEA Hamiltonian and optimal active spaces: CAS(4,6) and CAS(6,10) for CN3 and CN5, respectively.}
	\footnotetext[5]{CC3/ANO-L-VDZP excitation energies.}
	\footnotetext[6]{Diffusion Monte Carlo results based on optimal active space CASSCF trial wave functions obtained using the T$^\prime$+ basis set and a Jastrow factor including electron-nuclear and electron-electron terms.}
	\footnotetext[7]{Extrapolated CC3 excitation energies obtained by adding the difference between the CC3/ANO-L-VDZP and CC2/ANO-L-VDZP values to the CC2/ANO-L-VTZP results.}
\end{table}		

\subsection{
Multi-state example
\label{sec:MS-example}
}
We have chosen to illustrate the multi-state shifted-Bk algorithm presented in Sec.~\ref{sec:MS} by computing the first singlet transition energy of two cyanine dyes: CN3 (\ce{H2N-CH=NH2+}) and CN5 (\ce{H2N-CH=CH-CH=NH2+}).
This type of dyes are known to be particularly challenging for electronic structure methods, and especially time-dependent density-functional theory. \cite{Send_2011, Jacquemin_2012, Boulanger_2014, LeGuennic_2015}
The geometry of CN5 has been extracted from Ref.~\onlinecite{Jacquemin_2012} and we have optimized CN3 at the same level of theory (PBE0/cc-pVQZ).
Here, we use Dunning's aug-cc-pVDZ basis set which has been shown to be flexible enough to quantitatively model such transition thanks to the weak basis dependency of this valence $\pi \rightarrow \pi^\star$ transition. \cite{Send_2011, Loos_2018b}
In order to treat the two singlet electronic states on equal footing, a common set of determinants is used for both states.
In addition, state-averaged CASSCF(2,2) molecular orbitals, obtained with the GAMESS package, \cite{Schmidt_1993} are employed.

The difficulty of accurately modeling this vertical transition lies in the strong coupling between the $\sigma$ and $\pi$ spaces.
To assess this peculiar effect, we have performed several calculations and our results are gathered in Table \ref{tab:cya}.
(The corresponding total energies can be found in {\SI}.)
For comparison purposes, Table \ref{tab:cya} also reports reference calculations extracted from Ref.~\onlinecite{Send_2011}.
First, we have performed CAS-CI calculations taking into account only the set of molecular orbitals with $\pi$ symmetry.
We refer to these calculations as CAS($\pi$). 
For CN3 and CN5, there are, respectively, $4$ and $6$ electrons as well as $32$ and $50$ orbitals in the CAS($\pi$) space.
This results in multideterminant wave functions containing $11\,296$ and $670\,630$ determinants, respectively. 
To \alert{quantify} the strong coupling between the $\sigma$ and $\pi$ space, we have also computed full-valence exFCI energies [denoted as exFCI(\spp)]. \cite{Scemama_2018a, Loos_2018b} 
These values fits nicely with the exCC3(\spp) benchmark values reported by Send et al., \cite{Send_2011} in agreement with our previous study which shows that, at least for compact compounds, CC3 and exFCI yield similar excitation energies. \cite{Loos_2018b}

The difference between CAS($\pi$) and exFCI(\spp) is of the order of half an eV (slightly less for CN5), showing that the relaxation of the $\sigma$ orbitals plays a central role here, this effect becoming less pronounced when the number of carbon atoms increases.
Note that our CAS($\pi$) excitation energies are extremely close to the CASSCF results reported in Table \ref{tab:cya}.
The DMC estimates of Send et al.~\cite{Send_2011} are probably off by $0.2$ eV due to the lack of direct $\sigma$-$\pi$ coupling in the active space, which is only partially recovered by the Jastrow factor and the orbital optimization.

In CAS($\pi$)+\PT, the second-order correction $\ESS{2}{}$, computed by taking into account all the determinants from the FCI space connected to the CAS($\pi$) reference space, is added to the CAS($\pi$) result.
This correction goes in the right direction and recovers $0.19$ and $0.25$ eV for CN3 and CN5 respectively, bringing the excitation energies within $0.25$ and $0.13$ eV to the exFCI(\spp) values.

Similarly, CAS($\pi$)+{\sBkO} and CAS($\pi$)+{\sBk} correspond to {\sBk} and {\sBkO} calculations where the CAS($\pi$) model space is renormalized by the effect of the perturbers.
Like in the case of \ce{CuCl2}, CAS($\pi$)+{\sBkO} recovers slightly more than CAS($\pi$)+\PT, while CAS($\pi$)+{\sBk} is spot on for CN3, and overshoot slightly the exFCI(\spp) values for CN5 with an error of $0.12$ eV. 
These results shows that the shifted-Bk method associated with a CIPSI-like sCI algorithm is able to recover a large fraction of the missing correlation energy, even with relatively small model spaces.

\section*{Supplementary material}
See {\SI} for the pseudo-code of the multi-state algorithm, total energies associated with Table \ref{tab:cya} and exFCI extrapolations.

\begin{acknowledgments}
The authors would like to thank Jean-Paul Malrieu for stimulating discussions, \alert{and the anonymous referees for valuable comments and suggestions.}	
This work was performed using HPC resources from CALMIP (Toulouse) under allocations 2018-0510 and 2018-18005 and from GENCI-TGCC (Grant 2018-A0040801738).
\end{acknowledgments}

%

\end{document}